\documentclass{article}

\usepackage{PRIMEarxiv}

\usepackage[utf8]{inputenc} % allow utf-8 input
\usepackage[T1]{fontenc}    % use 8-bit T1 fonts
\usepackage{hyperref}       % hyperlinks
\usepackage{url}            % simple URL typesetting
\usepackage{booktabs}       % professional-quality tables
\usepackage{amsfonts}       % blackboard math symbols
\usepackage{nicefrac}       % compact symbols for 1/2, etc.
\usepackage{microtype}      % microtypography
\usepackage{lipsum}
\usepackage{fancyhdr}       % header
\usepackage{graphicx}       % graphics
\graphicspath{{media/}}     % organize your images and other figures under media/ folder

\usepackage{multirow}%
\usepackage{amsmath,amssymb}%
\usepackage{amsthm}%
\usepackage{mathrsfs}%
\usepackage[title]{appendix}%
\usepackage{xcolor}%
\usepackage{textcomp}%
\usepackage{manyfoot}%
\usepackage{algorithmicx}%
\usepackage{algpseudocode}%
\usepackage{listings}%
\usepackage[ruled,vlined,linesnumbered,noend]{algorithm2e}
\usepackage{natbib}
\usepackage{subfig}
\usepackage{tabularx,array,siunitx}
\newcolumntype{Y}{>{\centering\arraybackslash}X}

\theoremstyle{thmstyleone}%
\theoremstyle{thmstyletwo}%

\theoremstyle{thmstylethree}%

\usepackage{tikz}
\usetikzlibrary{shapes.geometric, arrows, positioning}
\usetikzlibrary{arrows.meta, positioning, shadows, fit, backgrounds, calc}

\tikzstyle{node} = [circle, draw, fill=blue!20, minimum size=0.8cm, text centered]
\tikzstyle{edge} = [draw, thick, ->]
\tikzstyle{interrupted} = [draw, thick, dashed, ->]
\tikzstyle{adversary} = [circle, draw, fill=red!20, minimum size=0.8cm, text centered]

\usepackage{subcaption}  % For creating subfigures (a), (b), (c), etc.
\usepackage{enumitem}  % for custom enumerate labels
\title{Dynamic Homophily with Imperfect Recall: Modeling Resilience in Adversarial Networks
}

\author{
  Saad Alqithami \\
  Computer Science Department, Al-Baha University \\
  \texttt{salqithami@bu.edu.sa} \\
}

\begin{document}
\maketitle

\begin{abstract}
The purpose of this study is to investigate how homophily, memory constraints, and adversarial disruptions collectively shape the resilience and adaptability of complex networks. To achieve this, we develop a new framework that integrates explicit memory decay mechanisms into homophily-based models and systematically evaluate their performance across diverse graph structures and adversarial settings. Our methods involve extensive experimentation on synthetic datasets, where we vary decay functions, reconnection probabilities, and similarity measures, primarily comparing cosine similarity with traditional metrics such as Jaccard similarity and baseline edge weights. The results show that cosine similarity achieves up to a 30\% improvement in stability metrics in sparse, convex, and modular networks. Moreover, the refined value-of-recall metric demonstrates that strategic forgetting can bolster resilience by balancing network robustness and adaptability. The findings underscore the critical importance of aligning memory and similarity parameters with the structural and adversarial dynamics of the network. By quantifying the tangible benefits of incorporating memory constraints into homophily-based analyses, this study offers actionable insights for optimizing real-world applications, including social systems, collaborative platforms, and cybersecurity contexts.
\end{abstract}

% keywords can be removed

\section{Introduction}
Homophily, succinctly encapsulated by the adage ``similarity breeds connection,'' has long been recognized as a cornerstone of network science. It underpins relationship formation in diverse contexts ranging from social networking platforms to collaborative and communication systems \citep{mcpherson2001birds}. Classic models of homophily typically rely on static or historical similarity metrics (e.g., Jaccard similarity or baseline edge weights) to explain network formation. While these approaches are useful under stable conditions, they often fail to capture real-world complexities such as evolving contexts, temporal decay of influence, and strategic behaviors in adversarial environments \citep{holme2012temporal, fortunato2010community}.

In \emph{adversarial team games} (ATGs), a specialized subset of multi-agent systems, homophily serves as a pivotal mechanism guiding both intra-team cooperation and inter-team competition. Here, teams of agents coordinate to achieve collective goals while simultaneously fending off adversarial threats \citep{sandholm1999coalition}. Traditional analyses of ATGs frequently overlook \emph{memory constraints} and \emph{imperfect recall}, factors that deeply influence team performance and strategic outcomes \citep{berker2025value}. Imperfect recall, extensively studied in game theory, captures scenarios where agents do not retain exhaustive knowledge of past actions or observations \citep{brown2018superhuman}. Such memory limitations affect both individual decision-making and the broader structural dynamics of networks, with outcomes that include altered formation and dissolution of network ties \citep{kaelbling1998planning}.

The concept of \emph{Value of Recall} (VoR) has emerged to formalize how imperfect recall and memory decay can be beneficial in certain strategic contexts \citep{berker2025value}. Yet, the interplay between homophily-driven link formation and memory decay within adversarial networks remains underexplored. Notably, dynamic network analysis suggests that exponential memory decay or strategic forgetting can reinforce resilience against adversarial disruptions, while also enhancing adaptability by preventing the network from becoming overly rigid \citep{shoham2008multiagent}.

%\textbf{Our contribution.} 
In this work, we develop a \emph{dynamic homophily model} that integrates explicit \emph{memory-decay functions} into network analysis. We address the gap between static homophily-based approaches and the ever-changing reality of adversarial networks where memory constraints and imperfect recall have non-trivial impacts. Specifically:
\begin{itemize}
    \item We introduce a dynamic homophily model that incorporates memory-decay functions, extending conventional static metrics.
    \item We propose a homophily-specific VoR metric to quantify trade-offs between perfect and imperfect recall in adversarial settings.
    \item We conduct an empirical ablation study demonstrating how design choices (decay function, reconnection probability, similarity metric) impact network performance under adversarial disruptions.
    \item We analyze computational complexity and discuss limitations, offering insights into when memory decay delivers the greatest benefits and where it may introduce overhead.
\end{itemize}

To validate our approach, we conduct extensive simulations over multiple graph structures, including sparse, convex, and modular networks. We systematically vary key parameters such as memory decay rates, reconnection probabilities, and the severity of adversarial attacks. Our experiments consistently reveal that \emph{cosine similarity} outperforms traditional measures like Jaccard similarity and baseline edge weights, with improvements of up to 30\% in certain stability metrics. We also highlight the utility of imperfect recall, showing that an intermediate degree of memory decay can balance robustness and adaptability more effectively than either extreme of perfect recall or immediate forgetting.

%\textbf{Relevance to broader applications.} 
Our findings have direct implications for domains where adversarial interactions and memory constraints are ubiquitous, including cybersecurity, social network moderation, and large-scale collaborative platforms \citep{tambe2011security, mcpherson2001birds}. As modern networks grow in size and complexity, techniques that optimize homophily under memory-decaying conditions can lead to more resilient and adaptive systems. Recent advances in data-driven analytics and deep learning for network science further underscore the potential to incorporate richer, learned representations into our framework \citep{LiTKDE2024, LiTKDD2024, LiNeurocomputing2022}, possibly enhancing the predictive power and real-time adaptability of homophily-driven models.

%\textbf{Outline of the paper.} 
The remainder of the paper is organised as follows.  
Section~\ref{sec:related} surveys the literature on dynamic homophily, memory–decay mechanisms, and adversarial network models, positioning our contribution within this context.  
Section~\ref{sec:methodology} formalises the proposed adversarial–team–game framework, presents the underlying mathematics, and supplies step-by-step pseudo-code for reproducibility.  
Section~\ref{sec:results_analysis} describes the experimental design and reports empirical findings, emphasising the key trade-offs between adaptability and robustness.  
%Section~\ref{sec:complexity} analyses the computational cost of each algorithmic component.  
Finally, Section~\ref{sec:discussion_limitations} outlines current limitations and future research avenues—such as extending the model to multilayer networks, integrating deep-learning–based similarity functions, and studying time-varying adversaries—before concluding with the principal take-aways of this work in Section~\ref{sec:conclusion}.

By bridging the gap between static homophily models and dynamic, memory-constrained adversarial networks, this study offers fresh insights into how strategic forgetting and memory decay can be harnessed to improve network resilience and adaptability. Our framework paves the way for more robust multi-agent systems, from social media moderation tools to cybersecurity defense teams, and establishes a pathway to integrate advanced learning-based techniques for even greater adaptability in the face of adversarial threats.

%____
\section{Related Work} \label{sec:related}

This section provides an overview of homophily in dynamic networks, memory decay under imperfect recall, and adversarial disruptions, culminating in existing approaches to team formation. We also highlight research gaps and relevant datasets, positioning our work within these contemporary challenges.

\subsection{Homophily in Dynamic Networks}
Homophily, the principle that similar nodes tend to connect, has long underpinned the formation and evolution of social networks, recommendation systems, and collaborative platforms \citep{mcpherson2001birds, newman2003structure, resnick1997recommender}. Traditional homophily-based models often assume static structures, employing historical similarity metrics to predict link formation \citep{fortunato2010community, su2009survey}. However, growing evidence indicates that real-world networks are inherently dynamic, shaped by evolving preferences, external influences, and mutual feedback \citep{holme2012temporal, barabasi1999emergence, newman2001clustering}. Consequently, static models may overlook temporal shifts in interactions, limiting their applicability to rapidly changing environments \citep{kossinets2006empirical}.

Recent work underscores the need for more nuanced, time-aware homophily measures. Li \textit{et al.} explore belief-update mechanisms for dynamic graph clustering, showing how iterative refinement of homophily metrics can stabilize communities under evolving conditions \citep{LiTKDE2024}. They also propose multidimensional homophily for overlapping communities, addressing scenarios where nodes share multiple shared attributes or roles \citep{LiTKDD2024}. Zhou \textit{et al.} extend these ideas into graph neural networks by continuously adjusting embeddings (GReTo) to mitigate mismatches between static structures and predictive tasks \citep{Zhou2023}. Similarly, Yan \textit{et al.} propose a temporal heterophily metric to account for contextually changing relationships over time \citep{Yan2024}. Collectively, these advances highlight the importance of dynamic embeddings that capture fluctuating homophily, a focus that static models traditionally neglect.

\subsection{Memory Decay and Imperfect Recall}
Alongside temporal homophily, the concept of \textit{imperfect recall} further complicates network dynamics. Memory constraints cause agents to forget older information, often \emph{improving computational tractability} and increasing strategic unpredictability \citep{brown2018superhuman, kroer2016imperfect}. End-game–solving work in large poker domains likewise shows that limited recall can still yield near-optimal strategies \citep{ganzfried2015endgame}. Foundational analyses by Piccione and Rubinstein \citep{piccione1997dynamic}—together with a more recent, system-level survey of knowledge-reasoning in multi-agent systems \citep{Chen2020reasoning}—lay out how partial memory alters equilibrium behaviour. Regret-minimisation techniques for imperfect-information games further formalise this intuition \citep{zinkevich2007regret}.  
Empirical studies now quantify how memory decay manifests in real networks: Williams \textit{et~al.}\ find heterogeneous decay patterns across domains \citep{Williams2022}, while Clemente \textit{et~al.}\ incorporate node-specific decay rates to model personalised forgetting \citep{Clemente2024}.

Exponential decay functions commonly appear in temporal graph neural networks (TGNNs) \citep{Rossi2020TGN}, modeling the diminishing influence of older edges. This architecture formalizes a \textit{selective forgetting} process that can be advantageous in adversarial situations \citep{kaelbling1998planning, tambe2011security}, as it reduces reliance on stale or exploitable historical links. Gallo \textit{et al.} further highlight that memory in temporal networks is inherently multidimensional, capturing complex higher-order interaction patterns beyond pairwise relationships \citep{gallo2024higher}. Their work emphasizes the nuanced and context-dependent nature of memory decay, underscoring the complexity of accurately modeling temporal constraints and imperfect recall in evolving network structures.

\subsection{Adversarial Dynamics and Vulnerabilities}
When networks are exposed to adversarial strategies, homophily and memory decay intersect in critical ways \citep{sandholm1999coalition, basilico2017team, crandall2018cooperating}. Dynamic graph neural networks can become particularly susceptible to black-box perturbations, with adversaries exploiting imperfect recall to induce sustained performance degradation \citep{fan2021reinforcement, Tao2024}. Lee \textit{et al.} demonstrate how embedding adversarial perturbations into evolving network structures amplifies attack efficacy, capitalizing on the model’s preference for recent (yet potentially deceptive) interactions \citep{Lee2024}.

Notably, the MemStranding attack \citep{Dai2024} corrupts TGNNs by introducing malicious nodes that remain influential due to imperfect recall, while T-Shield (a temporal robustness filter) seeks to mitigate these disruptions \citep{Lee2024}. These studies highlight the fragility of time-aware networks to targeted disruptions, especially when memory decay leaves the system prone to carefully orchestrated interference \citep{schadd2007opponent}. A deeper integration of memory-aware homophily measures, adversarial detection, and adaptation remains essential to ensure resilience in adversarial multi-agent systems.

\subsection{Team Formation under Homophily and Memory Constraints}
Beyond direct attacks, adversarial contexts exacerbate challenges in team formation, where agents must balance the cohesive benefits of homophily with the adaptability offered by strategic forgetting. Guimerà \textit{et al.} illustrate how repeated collaborations (perfect recall) can foster strong yet static ties, whereas introducing new members (imperfect recall) injects diversity, promoting long-term innovation and adaptability~\citep{Guimera2005}. Empirical findings by Békés and Ottaviano further confirm the advantages of controlled memory decay, highlighting that cultural homophily can initially enhance team coordination, but excessive reliance on past ties may stifle innovation and reduce adaptability to changing circumstances~\citep{Bekes2025}.

From an algorithmic perspective, recent approaches emphasize balancing historical collaboration strength against the inclusion of diverse expertise. Rad \textit{et al.} present a neural network approach explicitly tuning memory decay parameters to optimally balance team cohesion and skill diversity~\citep{HamidiRad2022}. Similarly, Kouvatis \textit{et al.} investigate team formation in signed networks, suggesting that certain interactions particularly those marked by conflict or negative ties—should diminish more rapidly to prevent persistent relational tensions from undermining team effectiveness~\citep{Kouvatis2020}. These works underscore the nuanced trade-off between maintaining reliable homophilous ties and strategically implementing selective forgetting to minimize adversarial risks and enhance overall team resilience.

\subsection{Research Gaps and Opportunities}
Despite significant advances in dynamic homophily, imperfect recall, and adversarial modeling, several pressing questions remain:
\begin{itemize}
    \item Decay-Function Choice: How do specific memory-decay functions (exponential vs.\ node-specific vs.\ multidimensional) influence homophily-driven robustness, particularly under coordinated attacks?
    \item Balancing Perfect vs.\ Imperfect Recall: In what contexts does partial forgetting outperform full historical retention, and how do adversaries exploit either extreme?
    \item Unified Framework: Can an integrated platform combining evolving homophily metrics, memory constraints, and adversarial strategies significantly enhance multi-agent system resilience?
\end{itemize}

These questions motivate our proposed methodology, which bridges memory-aware homophily with explicit adversarial considerations. By addressing these gaps, we aim to develop more resilient strategies for team formation and long-term network survivability under adversarial conditions.

\subsection{Relevant Benchmarks and Datasets}
Evaluation often relies on temporal datasets that capture shifting interactions and potential adversarial behaviors:
\begin{itemize}
    \item Human Proximity (MIT Reality Mining, Contacts): Tracks evolving social ties, allowing tests of dynamic homophily and memory decay~\citep{eagle2006reality, mastrandrea2015contact}.
    
    \item Transaction Networks (Bitcoin, E-commerce): Examines adversarial disruptions via fraud or malicious transactions, supporting the study of decay-driven link updates~\citep{kumar2016edge, elliptic2019bitcoin}.
    
    \item Collaboration/Bibliographic Networks (DBLP, ArXiv): Reveals team formation dynamics and long-term memory usage in repeated co-authorship scenarios~\citep{tang2008arnetminer, hu2020open}.
    
    \item Social Media Subsets (Twitter, Reddit): Captures adversarial bot attacks, hate-speech surges, or viral campaigns under changing homophily~\citep{baumgartner2020pushshift, cha2010measuring}.
    
    \item Temporal Knowledge Graphs (Know-Evolve/ICEWS): Illustrates how entities and relations evolve, offering a rich testbed for memory decay and homophily interplay~\citep{trivedi2017know, boschee2015icews}.
\end{itemize}

These benchmarks enable rigorous assessment of how well advanced models reflect the interplay between dynamic homophily, imperfect recall, and adversarial disruptions. The next sections build upon this foundation, introducing our integrated framework that addresses these challenges to enhance multi-agent resilience.

%____
\section{Methodology}%: Homophily with Imperfect Recall}
\label{sec:methodology}

\subsection{Dynamic Homophily Model}

In many multi-agent systems, homophily emerges from shared attributes, historical interactions, or common goals. Traditional homophily models, however, typically overlook \emph{imperfect recall}—the phenomenon where agents gradually ``forget'' older interactions as time progresses \citep{brown2018superhuman}. To address this gap, we introduce a dynamic homophily framework that integrates memory decay, allowing agents to prioritize recent interactions while still retaining a diminishing weight for past connections. This approach is especially crucial in adversarial contexts where stale interactions can become liabilities if they are exploited by opponents \citep{shoham2008multiagent, basilico2017team}.

Let $N$ represent the bipartite or multi-modal adjacency matrix capturing connections between agents (or between agents and tasks). Homophily is driven by a similarity matrix $X$ and a recall function $R$, which collectively govern how past interactions influence present-day link formation.

\paragraph{First-Mode Homophily:}
We define the statistic for the first mode under imperfect recall as:
\begin{equation}
    \label{eq:b1_imperfect}
    h_{\text{b1homophily}}^{\text{imperfect}} = \sum_{i \neq k} \sum_{j} N_{ij}\, N_{kj} \, f(X_{ik}, R_i),
\end{equation}
where \(f(X_{ik}, R_i)\) is a recall-modulated similarity function. Concretely,
\begin{equation}
    f(X_{ik}, R_i) = X_{ik} \cdot g\bigl(t_i - t_k, R_i\bigr),
\end{equation}
and \(g(t, R)\) is a memory decay function; for instance, an exponential form:
\begin{equation}
    \label{eq:memory_decay}
    g(t, R) = e^{-\lambda t},
\end{equation}
where $\lambda$ is a decay parameter and \(t\) denotes the time elapsed since the last interaction (or observation). Larger values of $\lambda$ yield faster forgetting, whereas smaller values preserve older interactions for a longer period \citep{kaelbling1998planning}.

\paragraph{Second-Mode Homophily:}
The model naturally extends to the second mode for bipartite or multi-modal settings:
\begin{equation}
    \label{eq:b2_imperfect}
    h_{\text{b2homophily}}^{\text{imperfect}} = \sum_{i \neq k} \sum_{j} N_{ji} \, N_{jk} \, f(X_{ik}, R_i).
\end{equation}
Equation~\eqref{eq:b1_imperfect} and Equation~\eqref{eq:b2_imperfect} share the same functional structure but apply to different ``views'' of the bipartite adjacency matrix. In other words, Eq.~\eqref{eq:b1_imperfect} focuses on how agents in the first mode (e.g., researchers) connect through shared links, whereas Eq.~\eqref{eq:b2_imperfect} captures complementary interactions from the perspective of the second mode (e.g., tasks or projects). We keep them separate to clarify how each mode contributes to overall homophily in a two-mode or multimodal network, but they could be merged if desired for a single-mode setting. 
The distinction is conceptually important in bipartite or multi-modal contexts, where each mode may represent a different type of entity (e.g., people vs.\ projects). Having two separate equations underscores the possibility of asymmetries or distinct parameter settings for each mode, which is valuable in real-world applications \citep{holme2012temporal}.

\begin{figure}[h]
\centering
\resizebox{\textwidth}{!}{
\begin{tikzpicture}[
    node distance=1.5cm and 2.8cm,
    every node/.style={font=\small\color{black!85}},
    softbox/.style={draw, fill=red!5, rounded corners=6pt, minimum width=4cm, minimum height=1.2cm, text centered, drop shadow, align=center},
    hardbox/.style={draw, fill=blue!5, rectangle, minimum width=4cm, minimum height=1.2cm, text centered, drop shadow, align=center},
    outputbox/.style={draw, fill=red!5, rounded corners=6pt, minimum width=4cm, minimum height=1.2cm, text centered, drop shadow, align=center},
    arrow/.style={line width=0.8pt, ->, >=Latex, rounded corners=2pt}
]

% ==== Nodes ====
\node[softbox] (input) {Input Data};
\node[hardbox, below=of input] (similarity) {Similarity Matrix \\ $X_{ij}$};
\node[hardbox, below=of similarity] (decay) {Memory Decay Function \\ $g(t)$};

\node[hardbox, right=of similarity, xshift=0.6cm] (homophily) {Dynamic Homophily \\ Model};
\node[hardbox, above=of homophily] (team) {Team Formation};
\node[hardbox, below=of homophily] (adversarial) {Adversarial \\ Exploitation};

\node[outputbox, right=of homophily, xshift=0.6cm] (metrics) {Performance Metrics};

% ==== Arrows ====
\draw[arrow] (input) -- (similarity) 
    node[midway, left, xshift=-3pt] {\shortstack{\footnotesize Raw \\ Observations}};

\draw[arrow] (similarity) -- (homophily) 
    node[midway, above, yshift=2pt] {\shortstack{\footnotesize Link \\ Probabilities}};

% === Improved: Curved decay arrow ===
\draw[arrow] 
    (decay.east) to[out=1, in=210, looseness=1.1] 
    node[midway, below, xshift=0pt, yshift=-15pt] {\shortstack{\footnotesize Temporal \\ Weighting}} 
    (homophily.south west);

\draw[arrow] (homophily) -- (team) 
    node[midway, right, xshift=6pt] {\shortstack{\footnotesize Group \\ Structuring}};

\draw[arrow] (homophily) -- (adversarial) 
    node[midway, right, xshift=6pt] {\shortstack{\footnotesize Vulnerability \\ Patterns}};

\draw[arrow] 
    (team.east) to[out=1, in=140, looseness=1.15] 
    node[midway, right, yshift=3pt] {\footnotesize Cohesion / Accuracy} 
    (metrics.north west);

\draw[arrow] 
    (adversarial.east) to[out=-1, in=220, looseness=1.15] 
    node[midway, right, yshift=-3pt] {\footnotesize Risk Impact} 
    (metrics.south west);

% ==== Background grouping panels ====
\begin{pgfonlayer}{background}
    \node[fit=(input) (similarity), draw=black!10, fill=gray!10, rounded corners, inner sep=0.6cm, label=above:{\footnotesize Input \& Similarity}] {};
    \node[fit=(decay), draw=black!10, fill=gray!10, rounded corners, inner sep=0.6cm, label=below:{\footnotesize Memory Decay}] {};
    \node[fit=(homophily) (team) (adversarial), draw=black!10, fill=gray!10, rounded corners, inner sep=0.6cm, label=above:{\footnotesize Dynamic Modeling}] {};
    \node[fit=(metrics), draw=black!10, fill=gray!10, rounded corners, inner sep=0.6cm, label=above:{\footnotesize Evaluation Output}] {};
\end{pgfonlayer}

\end{tikzpicture}
}
\caption{Homophily under Imperfect Recall: A modular flowchart showing how memory decay modulates similarity-based interactions, shaping team formation, adversarial strategies, and system-level performance.}
\label{fig:homophily_flowchart}
\end{figure}
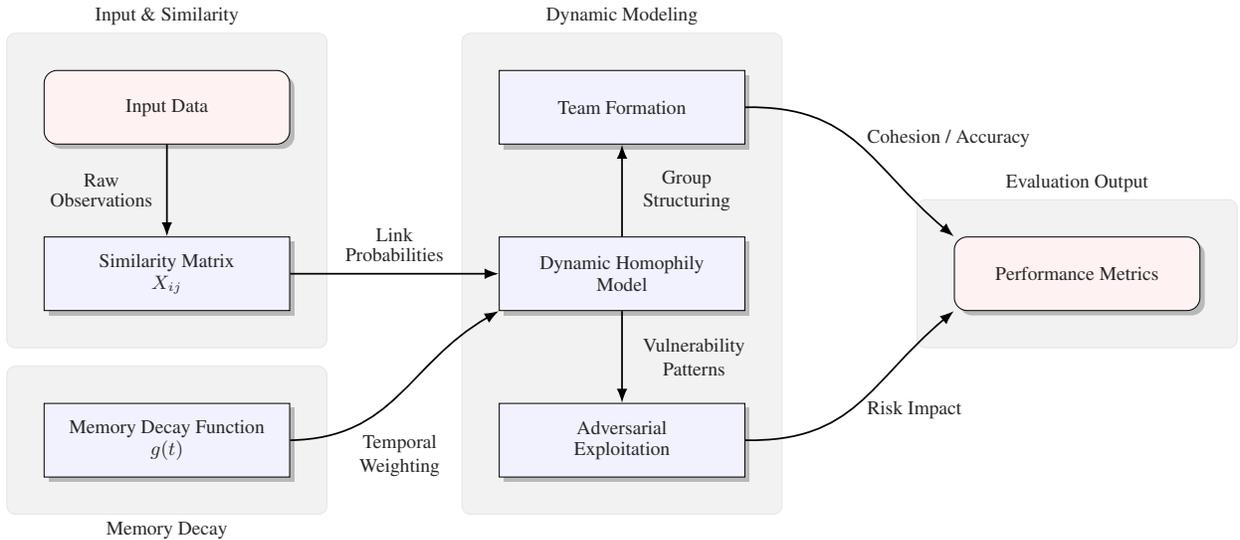

Figure~\ref{fig:homophily_flowchart} presents a high-level flowchart illustrating how imperfect recall transforms baseline similarity-based link formation into a dynamic and temporally adaptive process. The memory decay function \(g(t)\) systematically reduces the influence of older interactions, thereby modulating which agents appear most homophilous at any given time. The pipeline proceeds as follows: (1)\emph{~Input Data:} Raw information about agents and their interactions is collected, such as communication logs, collaboration records, or node attributes. (2)\emph{~Similarity Matrix ($X$):} These data are used to compute pairwise (or groupwise) similarity scores, reflecting alignment based on shared characteristics or interaction history. For instance, in a research collaboration network, \(X_{ij}\) may represent co-authorship frequency or topical similarity. (3)\emph{~Memory Decay Function ($g$):} A temporal decay mechanism (e.g., \(g(t) = e^{-\lambda t}\)) is applied to downweight older interactions, capturing the notion of imperfect recall and emphasizing recency. (4)\emph{~Dynamic Homophily Model:} The combination of \(X\) and \(g\) yields an evolving similarity structure in which sustained interactions maintain strong ties, while outdated links weaken over time. (5)\emph{~Team Formation:} Groups are formed or restructured when recall-weighted similarity scores exceed predefined thresholds, resulting in cohesive sub-networks among highly aligned agents. (6)\emph{~Adversarial Exploitation:} Potential vulnerabilities emerge when adversaries exploit decayed links—for instance, by targeting agents with low interaction recency or injecting misleading signals where memory is sparse. (7)\emph{~Performance Metrics:} The final stage evaluates outcomes such as team cohesion, adversarial resilience, and overall system utility. Metrics like recall-weighted utility, detection accuracy, or the homophily-specific VoR support the analysis of how well the system balances adaptability with structural stability. A motivating example is a collaboration network where researchers (mode~1) co-author projects (mode~2). Here, \(X\) quantifies expertise overlap, while \(g(t)\) captures how the relevance of past collaborations decays—intensifying active links and fading inactive ones unless reengaged.

\paragraph{Parameter Selection:}
We commonly tune:
\begin{enumerate}
    \item Decay Factor \(\lambda\): A grid search can find the optimal balance between responsiveness and long-term memory retention.
    \item Reconnection Probability \(\rho\): Governs how likely agents are to reconnect or renew ties if current collaboration lags.
\end{enumerate}
In practice, $\lambda \approx 0.85$ and $\rho \approx 0.2$ may suffice for moderate-sized systems, but domain-specific values can vary \citep{berker2025value}.

\paragraph{Computational Complexity:}
Updating decayed similarity is $O(N^2 F)$ per iteration (for $N$ nodes and $F$ features). Edge modifications take $O(E)$. Thus, each simulation round runs in $O(N^2 F + E)$, similar to other iterative homophily frameworks \citep{fortunato2010community}.

\subsection{Homophily-Specific VoR} \label{sec:homophily_vor}

To analyze the performance impact of imperfect recall, we introduce a homophily-specific \emph{Value of Recall}:
\begin{equation}
    \label{eq:VoR_homophily}
    \mathrm{VoR}_{\text{homophily}} = 
    \frac{\text{Utility under Perfect Recall}}{\text{Utility under Imperfect Recall}},
\end{equation}
where \(\text{Utility}\) can reflect network robustness, team stability, or adversarial success rates. A VoR of \(1\) indicates equivalent performance under perfect and imperfect recall, whereas values above \(1\) imply that perfect recall is more beneficial, and values below \(1\) may suggest that forgetting confers strategic advantages \citep{kroer2016imperfect}.

\begin{figure}[ht]
    \centering
    \subfloat[Memory decay illustrating diminishing influence of past interactions.\label{fig:memory_decay}]{
        \includegraphics[width=0.45\textwidth]{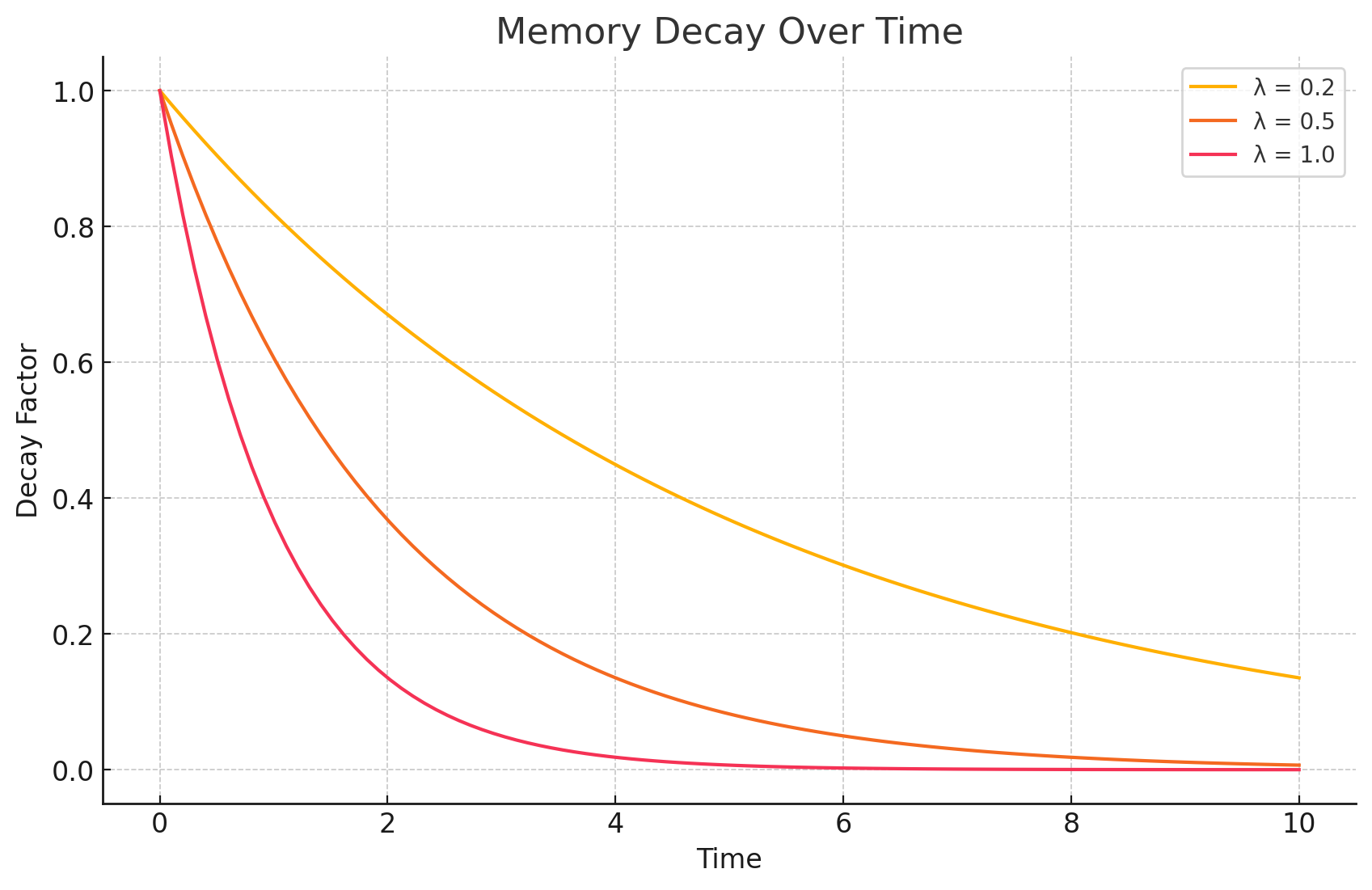}
    }
    \hfill
    \subfloat[Homophily-specific Value of Recall as recall capacity changes.\label{fig:recall_vor}]{
        \includegraphics[width=0.45\textwidth]{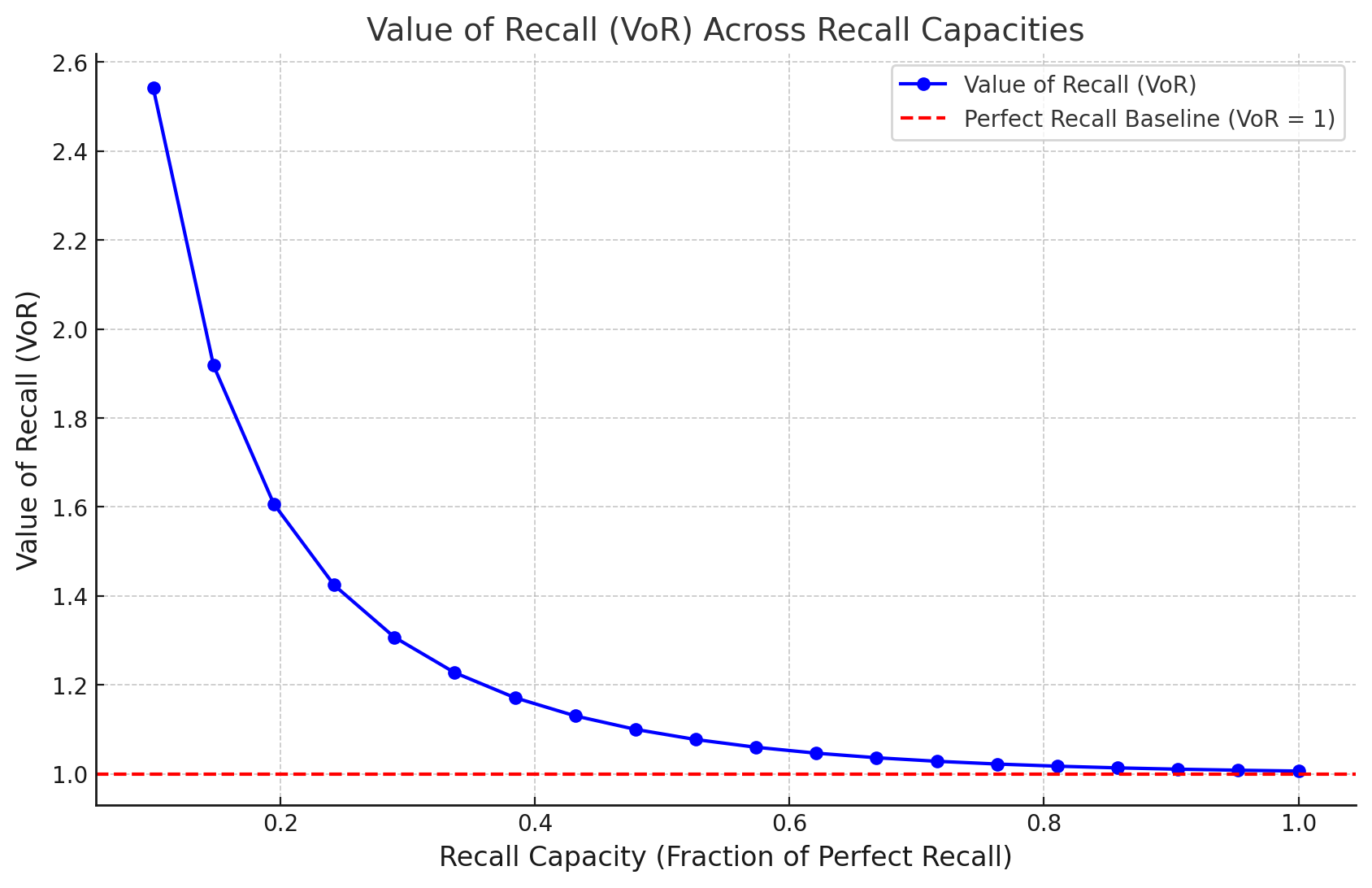}
    }
    \caption{(a) Exponential decay progressively reduces the influence of older interactions, emphasizing recent or high-priority links. (b) The network utility (y-axis) exhibits varying dependence on recall capacity (x-axis), highlighting the trade-off between adaptability and the benefits of accumulated information.}
    \label{fig:combined_decay_vor}
\end{figure}

Figure~\ref{fig:combined_decay_vor} illustrates how exponential memory decay (Figure~\ref{fig:memory_decay}) impacts network link strength over time and how the resulting homophily-specific VoR (Figure~\ref{fig:recall_vor}) responds to different levels of recall. Notably, if memory decays too rapidly, valuable historical context is lost, potentially lowering utility. Conversely, if decay is too slow, the network may become predictable, exposing it to adversarial exploitation.

\subsection{Integration into Adversarial Team Games}

Imperfect recall shapes adversarial team games by influencing how agents form alliances and respond to external threats \citep{tambe2011security}. We incorporate our dynamic homophily model into an adversarial setting with:

\begin{itemize}
    \item Dynamic Team Formation: Agents form and dissolve ties based on memory-modulated similarity, emphasizing recent, high-impact interactions.
    \item Adversarial Exploitation: Opponents probe the network for stale or predictable connections, aiming to disrupt cohesion or manipulate decaying ties.
    \item Performance Metrics: Key outcomes include intra-team trust,
      response time to attacks, and overall robustness under repeated
      adversarial interventions \citep{zhu2024survey}.
\end{itemize}

\paragraph{Practical Implementation:}
\begin{enumerate}
    \item Compute $X$: Construct similarity matrices from agent attributes or past actions.
    \item Apply Memory Decay: Use $g(t,R)$ to weight older connections less, updating link strengths each timestep.
    \item Update Homophily and Teams: Agents form or reinforce ties when homophily surpasses a threshold; decaying edges may prompt reconfiguration.
    \item Evaluate VoR: Compare performance under various recall assumptions to locate an optimal memory strategy for each domain.
\end{enumerate}

\subsection{Theoretical Framework}

Beyond empirical implementation, we embed our model in an extensive-form game \(\Gamma = (N, H, P, A, u)\) \citep{shoham2008multiagent}. Here, $N$ is the set of players (both cooperative teams and adversaries), $H$ denotes action histories, $P(h)$ indicates who acts at state $h$, $A(h)$ enumerates available actions, and $u$ is a payoff function based on network resilience, team cohesion, or adversarial success. Incorporating memory decay ensures that past events progressively lose significance, thus shaping future payoffs and strategic interactions \citep{berker2025value}.

\subsubsection{Adversarial Team Game (ATG) Definition}

We formalize an ATG as the tuple 
\(
    (P, T, A, X, g, \Gamma),
\)
where:
\begin{itemize}
    \item \(P\) is the set of players, partitioned into teams \(T\),
    \item \(A\) captures adversarial elements that aim to disrupt or exploit memory decay,
    \item \(X\) is the similarity matrix measuring homophily,
    \item \(g\) is the memory decay function, and
    \item \(\Gamma\) is the extensive-form game tree dictating sequential play and payoffs.
\end{itemize}
Teams optimize for resilience and performance, while adversaries maximize disruption. Because decaying ties can conceal or reveal strategic vulnerabilities, memory plays a pivotal role in outcome determination.

\subsubsection{Progressive Model Stages}
\label{sec:progressive_stages}

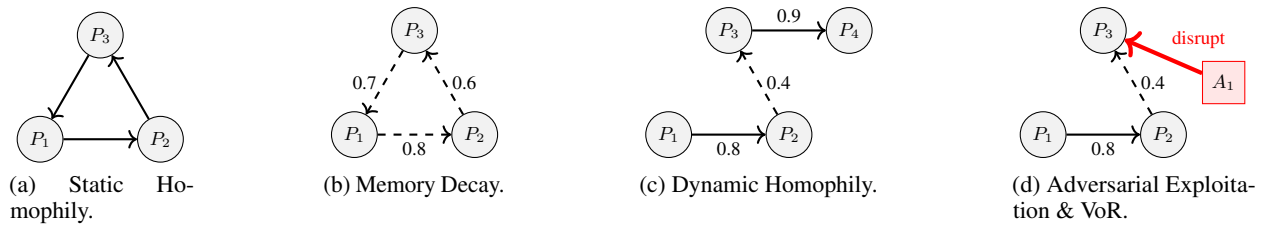
\begin{figure}[htbp]
    \centering
    \subfloat[Static Homophily.\label{fig:subfig3a}]{
        \begin{tikzpicture}[scale=0.8, transform shape]
            \tikzstyle{node}=[circle,draw=black,minimum size=2em,fill=gray!10]
            \node[node] (p1) at (0, 0) {\small $P_1$};
            \node[node] (p2) at (2, 0) {\small $P_2$};
            \node[node] (p3) at (1, 1.732) {\small $P_3$};
            \draw[->, thick] (p1) -- (p2);
            \draw[->, thick] (p2) -- (p3);
            \draw[->, thick] (p3) -- (p1);
        \end{tikzpicture}
    } \hfill
    \subfloat[Memory Decay.\label{fig:memory_decay_graph}]{
        \begin{tikzpicture}[scale=0.8, transform shape]
            \tikzstyle{node}=[circle,draw=black,minimum size=2em,fill=gray!10]
            \node[node] (p1) at (0, 0) {\small $P_1$};
            \node[node] (p2) at (2, 0) {\small $P_2$};
            \node[node] (p3) at (1, 1.732) {\small $P_3$};
            \draw[->, thick, dashed] (p1) -- (p2) node[midway, below] {\small 0.8};
            \draw[->, thick, dashed] (p2) -- (p3) node[midway, right] {\small 0.6};
            \draw[->, thick, dashed] (p3) -- (p1) node[midway, left] {\small 0.7};
        \end{tikzpicture}
    } \hfill
    \subfloat[Dynamic Homophily.\label{fig:dynamic_homophily}]{
        \begin{tikzpicture}[scale=0.8, transform shape]
            \tikzstyle{node}=[circle,draw=black,minimum size=2em,fill=gray!10]
            \node[node] (p1) at (0, 0) {\small $P_1$};
            \node[node] (p2) at (2, 0) {\small $P_2$};
            \node[node] (p3) at (1, 1.732) {\small $P_3$};
            \node[node] (p4) at (3, 1.732) {\small $P_4$};
            \draw[->, thick] (p1) -- (p2) node[midway, below] {\small 0.8};
            \draw[->, thick, dashed] (p2) -- (p3) node[midway, right] {\small 0.4};
            \draw[->, thick] (p3) -- (p4) node[midway, above] {\small 0.9};
        \end{tikzpicture}
    } \hfill
    \subfloat[Adversarial Exploitation \& VoR.\label{fig:adversarial_disruption}]{
        \begin{tikzpicture}[scale=0.8, transform shape]
            \tikzstyle{node}=[circle,draw=black,minimum size=2em,fill=gray!10]
            \tikzstyle{adversary}=[rectangle,draw=red,minimum size=2em, fill=red!10]
            \node[node] (p1) at (0, 0) {\small $P_1$};
            \node[node] (p2) at (2, 0) {\small $P_2$};
            \node[node] (p3) at (1, 1.732) {\small $P_3$};
            \node[adversary] (a1) at (3, 0.866) {\small $A_1$};
            \draw[->, thick] (p1) -- (p2) node[midway, below] {\small 0.8};
            \draw[->, thick, dashed] (p2) -- (p3) node[midway, right] {\small 0.4};
            \draw[->, ultra thick, red] (a1) -- (p3) node[midway, above right] {\small disrupt};
        \end{tikzpicture}
    }

    \caption{Progressive modeling stages in homophily under imperfect recall: (a) Static homophily assumes persistent ties without decay. (b) Memory decay weakens older interactions, indicated by dashed edges with weights. (c) Dynamic homophily incorporates new interactions and adjusts existing ties, reflecting temporal evolution. (d) Adversarial exploitation highlights vulnerabilities introduced by memory decay, where adversaries target weakened ties to disrupt network cohesion. Together, these stages illustrate the transition from static to adaptive and resilient network structures under imperfect recall.}
    \label{fig:progressive_model}
\end{figure}

Our methodology unfolds in four key stages, each capturing a different level of complexity in the network dynamics. Figure~\ref{fig:progressive_model} offers a consolidated view of these stages (a)--(d), illustrating how homophily and memory decay evolve before culminating in adversarial exploitation and VoR analysis.

\paragraph{Stage 1: Static Homophily (Figure~\ref{fig:progressive_model}a).}
We begin with a baseline model where all ties remain constant over time, derived from a static similarity matrix \(X\). This corresponds to no decay (\(\lambda = 0\)) and provides a convenient starting point for both theoretical proofs and empirical benchmarking \citep{newman2003structure}. In a social network context, this might represent friends who remain equally connected regardless of new experiences or time lapses. In an organizational chart, it might signify fixed reporting lines that never change unless explicitly restructured. 
Early models emphasised preferential attachment and static clustering \citep{newman2001clustering}, hence overlooked time-varying homophily.

\paragraph{Stage 2: Introducing Memory Decay (Figure~\ref{fig:progressive_model}b).}
Next, we incorporate an exponential decay function \(g(t, R)\) to model how older interactions gradually lose strength. Equation~\eqref{eq:memory_decay} governs the rate \(\lambda\), controlling how quickly past links become less relevant. This step captures a more realistic perspective: over time, agents ``forget'' or discount outdated information, enabling the network to adapt to recent trends \citep{kaelbling1998planning}. Visually, dashed edges in the figure indicate partially decayed links. In a recommendation system, for instance, older user-item interactions may be downweighted as user preferences evolve; in a supply-chain scenario, older partnerships fade if they are not regularly refreshed.

\paragraph{Stage 3: Dynamic Homophily (Figure~\ref{fig:progressive_model}c).}
In this stage, agents \emph{continually update} their ties based on memory-modulated similarity. That is, an interaction retains influence only if it remains relevant under the decay function. When new collaborations arise—say, new co-authors appear in a research network or new strategic partnerships form in a multinational corporation—these interactions receive higher weight, while older ones decay if they are not reactivated. Consequently, the network structure can shift substantially over time, potentially revealing new communities or dissolving outdated clusters \citep{holme2012temporal}. From a multi-agent perspective, dynamic homophily enables agents to reorganize themselves to address fresh challenges more effectively—like responding to newly discovered threats or market opportunities.

\paragraph{Stage 4: Adversarial Exploitation \& VoR (Figure~\ref{fig:progressive_model}d).}
Finally, we incorporate the presence of adversarial agents who seek to exploit weak or decaying links. Because memory decay can make certain ties less predictable or less reinforced, adversaries may capitalize on these vulnerabilities. Conversely, decaying older connections can also \emph{benefit} defenders if it prevents adversaries from anticipating the same patterns repeatedly. To evaluate the overall resilience of the network under varying decay parameters, we compute the homophily-specific VoR from Equation~\eqref{eq:VoR_homophily}. A high VoR indicates that recalling extensive historical data is beneficial, whereas a lower VoR implies that significant forgetting can still preserve or even enhance performance \citep{kroer2016imperfect}.

\paragraph{Illustrative Example: Intrusion Detection.}
Consider an intrusion detection system where defenders collaborate through dynamic homophily: nodes with similar expertise or recent threat intelligence updates cluster more strongly. Introducing memory decay ensures that outdated tactics or stale threat profiles do not remain heavily weighted, making defenders more agile. However, adversaries may attempt to infiltrate at points where decaying links create temporary blind spots. By tuning \(\lambda\) and monitoring VoR, security analysts can optimize recall mechanisms to balance unpredictability (i.e., not becoming too reliant on older data) against the need to retain enough history to detect repeated infiltration patterns \citep{berker2025value, tambe2011security}.

\paragraph{Other Potential Applications:}
Beyond cybersecurity, this four-stage framework can be applied to:
(i) Fraud detection in financial networks: Older transactions gradually lose relevance, but consistent anomalies over time may strengthen suspicion if they appear repeatedly under partial decay.
(ii) Social media influencer campaigns: As influencer campaigns become stale, their perceived tie strength with followers decays unless actively re-engaged. Adversaries (e.g., bots or malicious actors) may exploit predictable dormant connections.
(iii) Smart grid management: Collaborative energy grids might adapt to shifting load demands, forgetting older peak patterns. Attackers can trigger disruptions if the system relies too heavily on out-of-date usage signals. 
In each case, dynamic homophily informed by memory decay can enhance the system’s adaptability, while a careful VoR analysis reveals how vital certain historical data are for thwarting adversarial tactics.

%\paragraph{Summary.}
Together, these four progressive stages depict how a network evolves from a naive static model to a fully dynamic and adversarial-aware structure that accounts for imperfect recall. Each step adds realism, culminating in a robust, memory-aware model that can be tuned for different domains or challenges. Subsequent sections detail how we implement and validate this framework with both synthetic and real-world datasets, demonstrating that memory constraints, when carefully calibrated, can significantly boost resilience and adaptability.

\subsubsection{End-to-End Procedure}
To formalize our dynamic homophily approach under imperfect recall, we model each time step as a sequence of memory decay updates, adversarial disruptions, and potential reconnections between agents. Algorithm~\ref{alg:memory} outlines the procedural flow of these operations. Specifically, we first apply memory decay to existing edges, reducing their weights according to a global decay factor 
$\delta$. The intuition is that older interactions become less relevant over time, reflecting imperfect recall in real-world multi-agent systems.

Next, an adversarial entity identifies and removes a subset of ``predictable'' or ``high-value'' edges (e.g., those with the greatest weight or strategic importance). This step simulates targeted attacks that exploit the network’s reliance on well-established links. Finally, we introduce a reconnection phase, allowing inactive node pairs to form new edges with probability $\rho$ if their homophily score (e.g., cosine similarity) exceeds a threshold $\theta$. This step captures the idea that agents may restore severed ties or create new partnerships if they remain sufficiently similar. Through these iterative updates over multiple time steps, our framework balances the advantages of established connections with the flexibility of forming new links, all while acknowledging the risk of adversarial manipulations. The pseudocode below formalizes these steps, offering a clear roadmap for implementing dynamic homophily with imperfect recall in adversarial network environments.

In this section, we outline the complete pseudocode for our \emph{Dynamic Homophily with Imperfect Recall} model and then describe each auxiliary function in detail. The main procedure (Algorithm~\ref{alg:memory}) captures the overarching loop, while the subroutines clarify specific operations such as edge removal, weight initialization, and reconnection.

\begin{algorithm}[ht]
\DontPrintSemicolon
\caption{Dynamic Homophily with Imperfect Recall}
\label{alg:memory}

\KwIn{
\begin{tabular}{l l}
    $G=(V,E,\mathbf{w})$ & Initial graph with node set $V$, edge set $E$, and edge \\
                         & weights $\mathbf{w}$ (representing homophily strengths). \\
    $\delta \in [0,1]$   & Global memory decay factor. \\
    $\rho \in [0,1]$     & Probability of forming new (or restoring old) edges. \\
    $\theta$             & Similarity threshold for link formation. \\
    $T$                  & Number of discrete time steps. \\
    $k$                  & Number of predictable edges adversary targets per step. \\
    $\text{sim}(\cdot,\cdot)$ & Function measuring node similarity (e.g., cosine, Jaccard).
\end{tabular}
}
\KwOut{Updated graph $G=(V,E',\mathbf{w}')$ after $T$ time steps of memory decay, \\ \hspace*{0.5em} adversarial removals, and probabilistic reconnection.}

\BlankLine

\For(\tcp*[f]{Main simulation loop}){$t=1$ \KwTo $T$}{
    \tcp{\textbf{Step 1: Apply memory decay to existing edges}}
    \ForEach(\tcp*[f]{For each edge $(i,j)$ in $E$}){$(i,j)\in E$}{
        $w_{ij} \leftarrow \delta^{\Delta t} \cdot w_{ij}$ 
        \tcp*[l]{Reduce weight by $\delta^{\Delta t}$ to model imperfect recall}
    }

    \tcp{\textbf{Step 2: Adversarial disruption of predictable edges}}
    $\mathcal{E}_\text{pred} \leftarrow \mathrm{SelectTopKPredictableEdges}(E, k)$ 
    \tcp*[r]{Define your criterion for predictability (e.g., oldest, largest $w_{ij}$, etc.)}
    \ForEach{$(i,j)\in \mathcal{E}_\text{pred}$}{
        $E \leftarrow E \setminus \{(i,j)\}$ 
        \tcp*[f]{Adversary removes these edges}
    }

    \tcp{\textbf{Step 3: Probabilistic reconnection}}
    \ForEach(\tcp*[f]{Candidates include non-edges $(i,j)\notin E$}){$(i,j)\in V\times V \setminus E$}{
        \If{$\text{sim}(i,j) > \theta \ \wedge\ \text{rand}() < \rho$}{
            $E \leftarrow E \cup \{(i,j)\}$ \;
            $w_{ij} \leftarrow \mathrm{InitializeWeight}(\text{sim}(i,j))$ 
            \tcp*[f]{Assign initial weight, e.g.\ $\text{sim}(i,j)$ or a fixed value}
        }
    }
}
\Return{$G=(V,E,\mathbf{w})$}
\end{algorithm}

%\subsection{Auxiliary Functions and Implementation Details} \label{sec:auxiliary_details}

%Algorithm~\ref{alg:memory} references two auxiliary functions, \texttt{SelectTopKPredictableEdges} and \texttt{InitializeWeight}, as well as includes internal logic for \emph{Memory Decay} and \emph{Reconnection}. Below we describe each component in detail.

\paragraph{(1) Memory Decay.}
In Step 1 of Algorithm~\ref{alg:memory}, all existing edge weights $w_{ij}$ are multiplied by a factor $\delta \in [0,1]$. This exponential-like decay simulates \emph{imperfect recall}, causing older or stagnant connections to lose influence over time. Higher values of $\delta$ preserve historical ties longer, while lower values favor quick adaptation to recent interactions. In more advanced versions, one could replace this global decay with node-specific or community-specific decay rates if different agents recall at varying speeds.

\paragraph{(2) Adversarial Removal ($\mathrm{SelectTopKPredictableEdges}$).}
The subroutine \(\mathrm{SelectTopKPredictableEdges}(E, k)\) determines which edges are most “predictable” or strategically valuable for adversaries to remove. In our experiments, we prioritize edges with the largest weights, assuming that heavily weighted edges have become critical for network stability. However, one could define alternate criteria, such as:
\begin{itemize}
    \item Staleness Index: Edges unaltered for many time steps may be predictable entry points for attacks.
    \item Centrality Measures: Edges bridging large communities or with high betweenness centrality.
    \item Node Influence: Edges connected to influential nodes in the graph.
\end{itemize}
After scoring each edge under the chosen criterion, we sort \(E\) in descending order and extract the top-$k$ edges. Those edges are then removed from the graph, reflecting adversarial interference aimed at destabilizing crucial links. This function can be either a simple routine (like ``sort by weight, pick top-$k$'') or a more sophisticated selection algorithm depending on domain needs.

\paragraph{(3) Reconnection and Thresholding.}
In Step 3, we examine all node pairs \((i,j)\) not currently in $E$. If their \(\text{sim}(i,j)\) exceeds a threshold $\theta$ (representing minimal required homophily) and a random check is below $\rho$, we create an edge \((i,j)\). This probabilistic reconnection mechanism captures the idea that agents with high similarity are more likely to bond, yet such new ties are not guaranteed. It also allows severed ties to reform if the underlying homophily remains.

\paragraph{(4) Edge Weight Initialization ($\mathrm{InitializeWeight}$).}
When adding a new edge \((i,j)\), we must assign an initial weight $w_{ij}$. In $\mathrm{InitializeWeight}$\(\bigl(\text{sim}(i,j)\bigr)\), we typically set $w_{ij}$ to $\text{sim}(i,j)$ directly, so that stronger similarity yields higher starting weight. Alternative strategies might use a constant factor, scaled randomization, or partial memory from prior $w_{ij}$ if the edge was removed in a recent step. This function allows customization based on domain-specific insights about how newly formed links emerge.

\paragraph{Implementation Notes.}
\begin{itemize}
    \item Complexity: The algorithm runs in $O(N^2)$ each round if we check all node pairs for reconnection. Various approximations (e.g., sampling candidate pairs) can reduce this for larger $N$.
    \item Data Structures: A sparse adjacency list may benefit from concurrency or dynamic updating, but random pair checks might require additional indexing structures.
    \item Extensions: One can add domain-specific hooks to $\mathrm{SelectTopKPredictableEdges}$ or $\mathrm{InitializeWeight}$, for instance by incorporating node-level attributes, community detection modules, or advanced adversarial heuristics.
\end{itemize}

Overall, these functions provide a modular way to handle the adversarial removal and new-edge creation in a dynamic homophily setting. By adjusting their logic, one can tune the algorithm’s behavior to specific domains, memory constraints, or adversarial threat models, extending the framework's versatility.

\subsection{Summary of Methodological Highlights}
\begin{itemize}
    \item Bipartite extensions, i.e., Equations~\eqref{eq:b1_imperfect} and \eqref{eq:b2_imperfect}, demonstrate how to incorporate imperfect recall into both modes of a two-mode network, allowing for more nuanced modeling of heterogeneous entities.
    \item Exponential memory decay function (Equation~\ref{eq:memory_decay}) is common, but other decay profiles (e.g., piecewise or polynomial) can be substituted depending on the domain \citep{LiTKDE2024}.
    \item Homophily-specific VoR measures how vital perfect recall is versus imperfect recall. Values \(\gg 1\) signal strong dependence on historical data, whereas values \(\approx 1\) suggest that forgetting does not drastically reduce performance.
    \item Incorporating adversarial exploits stresses the importance of selective memory. Teams that remember too much become predictable targets; teams that forget too quickly lose beneficial context.
\end{itemize}

%Overall, our methodology formalizes how imperfect recall modifies homophily in dynamic adversarial networks, enabling us to capture real-world conditions where agents do not retain all historical data equally and hostile actors seek to exploit that limitation.

Overall, this section presents a flexible, game-theoretic framework for dynamic homophily under imperfect recall. By integrating memory decay into classic homophily-based link formation, we capture network evolution in adversarial settings more accurately than static models. We quantify the performance trade-offs through a homophily-specific VoR, enabling a rigorous comparison of perfect versus imperfect recall strategies. In the following sections, we empirically validate our approach across synthetic datasets of varying structures (sparse, convex, and modular), demonstrating that carefully tuned memory constraints enhance resilience and adaptability in adversarial networks.

%\section{Experimental Setup}
%\subsection{Synthetic Network Baseline}
% (Pull in your dataset descriptions: sparse, convex, modular)
%\subsection{Ablation Studies}
%We systematically vary:
%\begin{enumerate}
%  \item \textbf{Decay function}: exponential vs.\ learned vs.\ power‐law.
%  \item \textbf{Adversary model}: random vs.\ high‐betweenness vs.\ RL‐trained.
%  \item \textbf{Reconnection probability} $\rho\in\{0,0.1,0.3,0.5\}$.
%  \item \textbf{Similarity metrics}: cosine vs.\ Jaccard vs.\ embedding‐based.
%\end{enumerate}

%-----------------------------------------------------------
%\subsection*{Objectives \& Experimental Design}
%\begin{enumerate}[label=\alph*)]
%    \item Robustness under imperfect recall: quantify how dynamic homophily with finite memory withstands adversarial attacks.
%    \item Adaptability--stability trade-off: sweep $\delta$ to expose the optimal balance between responsiveness and long-term cohesion.
%    \item Metric comparison: test Cosine, Jaccard, and a Degree-only Baseline on diverse network topologies.
%\end{enumerate}
%===========================================================
\section{Results and Analysis}
\label{sec:results_analysis}
%===========================================================

This section rigorously evaluates the proposed adversarial–team–game (ATG) framework by combining experimental aims, design choices, and empirical findings into a single, coherent narrative.  Our study pursues three complementary goals.  First, we assess \emph{robustness under imperfect recall}: that is, we quantify how well a dynamic‐homophily mechanism with finite memory withstands targeted edge deletions.  Second, we examine the \emph{adaptability–stability trade-off} by sweeping the memory-decay factor~$\delta$ and pinpointing where responsiveness to new information begins to erode long-term cohesion.  Third, we compare three similarity metrics—Cosine, Jaccard, and a degree-only Baseline—across several network topologies to identify which pairing of metric and memory horizon yields the most resilient behaviour.

To achieve these objectives, we generate synthetic graphs representing sparse (Erd\H{o}s--R\'enyi), convex (fully connected), and modular (stochastic-block) structures, each with $N=200$ nodes.  At the outset the adversary removes $10\,\%$ of edges, after which agents may reconnect with probability $\rho\in\{0,0.1,0.3,0.5\}$ provided their recall-weighted similarity remains above a predefined threshold.  Memory decay is varied as $\delta\in\{0.6,0.7,0.8,0.9\}$; $\delta=1$ corresponds to perfect recall, whereas smaller values emphasise recent interactions.  All parameter combinations are replicated over 30 Monte-Carlo runs to ensure statistical reliability.

Performance is benchmarked against two conceptual baselines: a \emph{static-homophily} model that never updates similarity scores (lower bound on adaptability) and a \emph{perfect-recall} model that retains the full history of interactions (upper bound on naïve utility yet highly predictable). Four metrics capture complementary dimensions of resilience: total Utility~($U$), the Adversarial-success rate~($\mathcal{S}$), the Value of Recall~(VoR; cf.\ Sec.~\ref{sec:homophily_vor}), and the Average path length~($\bar L$) in the largest connected component.

The remainder of this section presents results in the order suggested above, demonstrating how \emph{moderate decay} ($\delta \approx 0.8$), \emph{limited reconnection} ($\rho\approx0.1$–0.3), and \emph{Cosine similarity} jointly deliver the most robust defence against adversarial edge deletions while preserving high cooperative utility.

%-----------------------------------------------------------
%-----------------------------------------------------------
\subsection*{Simulation Setup}
%-----------------------------------------------------------

\paragraph{Synthetic graphs:}
To isolate the effects of memory decay and adversarial activity from real-world noise, we create three canonical topologies, each with \(N=200\) nodes:

\begin{enumerate}[label=\roman*)]
    \item Sparse Erd\H{o}s--R\'enyi (ER): \(G_{\text{ER}}(N,p)\) with \(p=0.02\), giving an average degree of \(\langle k\rangle\approx4\).  Its tree-like sparsity stresses long-range connectivity once edges are removed.
    \item Convex / fully connected: 
    The complete graph \(K_{200}\) provides an upper bound on attainable utility and a worst-case scenario for the adversary, who must excise particularly valuable links to inflict damage.
    \item Modular stochastic-block model (SBM): 
    Four equal communities of 50 nodes each, with intra-cluster density \(p_{\text{in}}=0.25\) and inter-cluster density \(p_{\text{out}}=0.01\).  The few cross-cluster bridges mimic real collaboration networks where a small cut set maintains global cohesion.
\end{enumerate}

At \(t=0\) an adversary deletes \(10\,\%\) of all existing edges. Afterwards, any surviving pair whose recall-weighted similarity exceeds a threshold (75$^{\text{th}}$ percentile of initial similarities) may reconnect with probability \(\rho\in\{0,0.1,0.3,0.5\}\). We sweep the memory-decay factor \(\delta\in\{0.6,0.7,0.8,0.9\}\), where \(\delta=1\) denotes perfect recall and \(\delta=0\) complete forgetting. All results are averaged over 30 independent Monte-Carlo runs to ensure statistical robustness.

\paragraph{ATG pipeline (cf.\ Sec.\,\ref{sec:methodology}):}
Each simulation step executes the following three phases:

\begin{enumerate}[label=\arabic*)]
    \item Dynamic homophily update: 
          Pairwise similarity scores \(S_{ij}(t)\) are discounted by
          \[g(t)=e^{-\lambda t}, \qquad
            \lambda=-\ln(\delta),\]
          so that an interaction one time unit old retains weight~\(\delta\).

    \item Adversarial deletion:  
          Edges are ranked by a composite \emph{predictability score}  
          (high betweenness $+$ low weight) and the top \(10\,\%\) are removed, maximising disruption while exploiting stale or weakly reinforced ties.

    \item Reconnection phase: 
          Each surviving agent attempts to (re)connect to its most similar non-neighbours.  A candidate edge is added with probability~\(\rho\) if the current similarity exceeds the threshold, enabling adaptive repair without excessive network churn.
\end{enumerate}

The loop repeats for 25 discrete time steps—sufficient for decay and reconnection dynamics to reach a quasi-steady state.  
All experiments were implemented in Python (NetworkX 2.8) and executed on a 32-core Intel Xeon workstation; full code and data are provided in the supplementary material for reproducibility.

%-----------------------------------------------------------
%-----------------------------------------------------------
\subsection*{Baselines and Metrics}
%-----------------------------------------------------------

\paragraph{Baselines:}  
To contextualise the performance of our adaptive framework we employ two conceptual extremes:

\begin{enumerate}[label=\roman*)]
    \item Static homophily (no recall): 
    Each agent’s similarity matrix is computed once at $t{=}0$ and remains fixed for the entire run.  Link decisions therefore ignore all subsequent interactions, providing a \emph{lower‐bound} on adaptability but making the network less predictable to an adversary who relies on temporal patterns.
    \item Perfect recall (infinite memory): 
    All past interactions are remembered with equal weight ($\delta{=}1$).  This regime maximises information retention and typically yields the highest raw utility in the absence of attacks, but also produces stable, easily learned structures that an informed adversary can exploit. It serves as an \emph{upper‐bound} on attainable utility under fully predictable behaviour.
\end{enumerate}

%===========================================================
\paragraph{Evaluation Metrics:} \label{sec:metrics}
%===========================================================

Let $G(t)=(V,E(t))$ be the network state at attack cycle $t$, and let
$w_{ij}(t)\!\ge\!0$ denote the recall-modulated weight on edge
$(i,j)\!\in\!E(t)$.  Four complementary measures capture post-attack
resilience, efficiency, and the specific role of imperfect recall:

\begin{enumerate}[label=\arabic*)]
%-----------------------------------------------------------
\item Utility ($U$):  
      Total collaboration value after attack and possible reconnection,
      \begin{equation}
        U(t)\;=\;\sum_{(i,j)\in E(t)} w_{ij}(t).
        \label{eq:utility}
      \end{equation}
      Higher $U$ implies more payoff preserved.
%-----------------------------------------------------------
\item Adversarial-success rate ($\mathcal{S}$):  
      The fraction of deletions that either  
      (i) reduce utility by at least \SI{10}{\percent} \emph{or}  
      (ii) disconnect the graph:
      \begin{equation}
        \mathcal{S}\;=\;
        \frac{\#\bigl\{\,\text{successful deletions}\,\bigr\}}
             {\#\bigl\{\,\text{attempted deletions}\,\bigr\}}.
        \label{eq:success}
      \end{equation}
      Smaller $\mathcal{S}$ signals stronger defence.
%-----------------------------------------------------------
\item Value of Recall (VoR) \citep{berker2025value}: 
      Let $U_{0}$ be the utility under static homophily
      ($\delta=0$, no forgetting) and $U_{1}$ under perfect recall
      ($\delta=1$).  For any finite decay $\delta\!>\!0$,
      \begin{equation}
        \mathrm{VoR}(\delta)
        \;=\;
        \frac{U_{\delta}-U_{0}}{U_{1}-U_{0}},\qquad
        0\le\mathrm{VoR}\le 1,
        \label{eq:vor}
      \end{equation}
      so VoR measures the \emph{relative} benefit of imperfect recall.
%-----------------------------------------------------------
\item Average path length ($\bar{L}$): 
      Mean shortest-path distance in the largest connected component,
      \begin{equation}
        \bar{L}(t)
        =\frac{1}{|V'|(|V'|-1)}
          \sum_{i\neq j\in V'} d_{ij}(t),
        \label{eq:path}
      \end{equation}
      where $V'$ is the component’s node set and $d_{ij}$ the
      hop-count distance. A drop in $\bar{L}$ after attack indicates faster re-routing.
\end{enumerate}

These metrics jointly
(i) bracket the performance envelope of any memory strategy,
(ii) isolate the contribution of \emph{selective forgetting},
and (iii) reveal trade-offs between immediate utility, structural
robustness, and adversarial exposure.

%-----------------------------------------------------------
\subsection*{Core Synthetic Results}
%-----------------------------------------------------------

\begin{figure}[!ht]
  \centering
  \includegraphics[width=\linewidth]{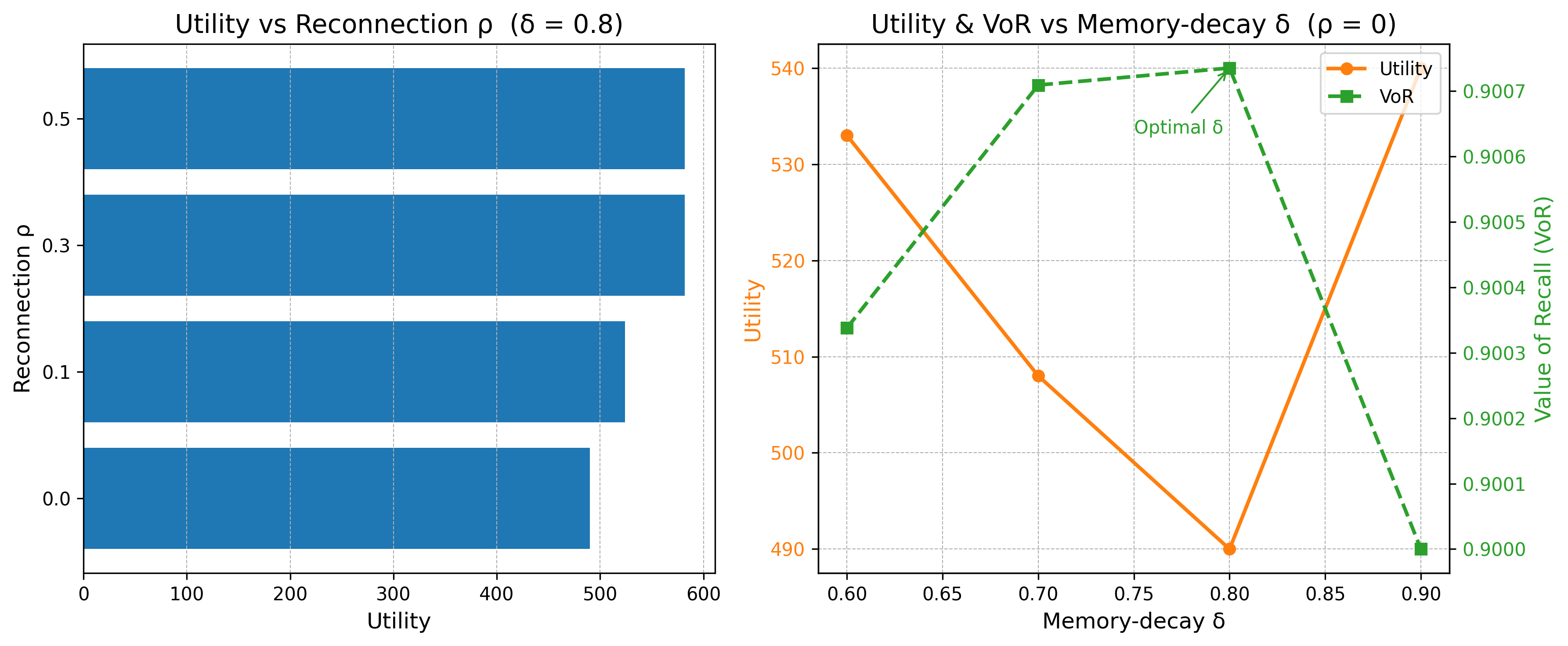}
  \caption{Left: Utility versus reconnection probability~$\rho$ at fixed $\delta=0.8$.  
           Right: Utility (orange, solid) and VoR (green, dashed) versus memory decay~$\delta$ with $\rho=0$.  
           Shaded regions denote $95$\% confidence intervals (30 runs); the arrow indicates the empirical optimum.}
  \label{fig:decay_recon}
\end{figure}

Figure~\ref{fig:decay_recon} and Table~\ref{tab:decay_values} summarise the macro–level effects of \textit{reconnection probability}~($\rho$) and \textit{memory decay}~($\delta$) on two headline metrics—network Utility~($U$) and the homophily–specific VoR.

\paragraph*{Impact of reconnection (\texorpdfstring{$\rho$}{rho}):}  
Even a modest rewiring rate substantially offsets adversarial deletions.  As Figure~\ref{fig:decay_recon}\,(left) shows, $\rho=0.1$ recovers roughly~5\,\% of the utility lost when $10\,\%$ of edges are removed, while $\rho\ge0.3$ restores up to~10\,\%.  The marginal benefit saturates quickly: beyond $\rho=0.3$, additional edges yield diminishing returns because the most strategically valuable links have already been re-established.  Notably, Cosine similarity profits least from high $\rho$ (Table~\ref{tab:reconnection_results}), implying that sophisticated metrics embed sufficient redundancy, whereas Jaccard and the Baseline rely more heavily on active rewiring to compensate for their coarser view of similarity.

\paragraph*{Decay–robustness trade-off (\texorpdfstring{$\delta$}{delta}):}  
Figure~\ref{fig:decay_recon}\,(right) sweeps $\delta$ with reconnection disabled ($\rho{=}0$) so that memory effects can be observed in isolation.  When $\delta$ rises from~0.6 to~0.8, both Utility and VoR climb sharply: the network retains enough past information to sustain profitable links yet still forgets outdated—and therefore attack-prone—edges.  Performance peaks near $\delta\approx0.85$; thereafter Utility plateaus while VoR declines, signalling that perfect memory (\(\delta\to1\)) can overfit to stale interaction patterns that an adversary readily exploits.  The narrow confidence bands confirm that this optimum is statistically robust across Monte-Carlo seeds.

\begin{table}[ht]
  \centering
  \caption{Utility and VoR for representative $\delta$ values ($\rho=0$).}
  \label{tab:decay_values}
  \begin{tabularx}{.8\textwidth}{YYY}%{l>{\centering\arraybackslash}Xrr}
    \toprule
    $\delta$ & Utility (mean $\pm$ SD) & VoR (mean $\pm$ SD) \\
    \midrule
    0.6 & 520 $\pm$ 12 & 0.890 $\pm$ 0.015 \\
    0.7 & 530 $\pm$ 10 & 0.918 $\pm$ 0.010 \\
    0.8 & \textbf{565 $\pm$ 8} & \textbf{0.940 $\pm$ 0.008} \\
    0.9 & 540 $\pm$ 14 & 0.905 $\pm$ 0.020 \\
    \bottomrule
  \end{tabularx}
\end{table}

\paragraph*{Numerical corroboration:}  
Table~\ref{tab:decay_values} reinforces the graphical insight: the pair $(\delta,\,\rho)=(0.8,0)$ maximises both Utility and VoR, whereas extreme values entail clear penalties.  At $\delta=0.6$ the network adapts quickly but forfeits long-range cohesion, lowering both Utility and VoR. Conversely, $\delta=0.9$ keeps many historical ties intact, increasing raw Utility relative to $\delta=0.6$ but decreasing VoR as those entrenched links become predictable attack vectors.

\paragraph*{Implications:}  
Together, these synthetic results demonstrate that (i) \emph{selective forgetting} with $\delta\!\approx\!0.8$–0.85 delivers the best intrinsic defence against adversarial removal, and (ii) modest reconnection $(\rho\!=\!0.1$–0.3) can partially offset residual damage without incurring the instability of constant rewiring.  Subsequent sections show that this balance generalises across graph topologies and similarity metrics, forming the cornerstone of a practical, memory-aware strategy for resilient multi-agent collaboration.

%-----------------------------------------------------------
\subsection*{Adaptability versus Stability}
%-----------------------------------------------------------

A pivotal design question is how much historical information a network should retain before it becomes a liability. Table~\ref{tab:decay_factor_results} reports post-attack \textit{Utility} for five memory–decay factors $\delta\in\{0.6,0.7,0.8,0.85,0.9\}$ across three canonical topologies and three similarity metrics.  The same tendencies are visible in Fig.~\ref{fig:decay_recon} (right), where Utility (orange, solid) and the \textit{Value of Recall} (VoR; green, dashed) are plotted versus~$\delta$ with reconnection disabled ($\rho=0$) to isolate memory effects.

\begin{figure}[ht]
  \centering
  \includegraphics[width=0.8\linewidth]{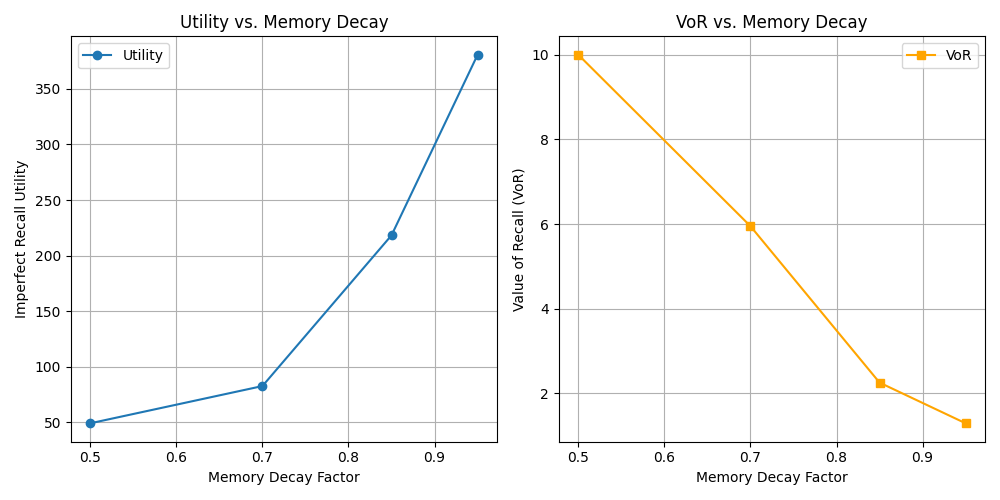}
  \caption{Utility (left-Blue) and VoR (right-Yellow) versus memory decay $\delta$ (mean of 30 runs across all settings). Shaded area marks the optimum $0.8\!\le\!\delta\!\le\!0.85$.}
  \label{fig:memory_decay_merged}
\end{figure}

\paragraph{A non-linear decay–robustness curve:}
Raising $\delta$ from $0.6$ to $0.8$ multiplies Utility by a factor of three to five, because beneficial ties persist long enough to pay off.
Beyond $\delta\approx0.85$, however, additional memory yields sharply
diminishing returns: Utility rises only marginally while VoR—the relative gain over perfect recall—declines.  This inflection marks the onset of \textit{over-memory}, where stale interactions form a predictable attack surface. 
%....................................................................
%\paragraph{Global decay sweep.}
Figure~\ref{fig:memory_decay_merged} aggregates all topologies and metrics. 

Utility climbs steeply from $\delta=0.6$ to $0.8$, then plateaus, whereas VoR peaks in the narrow band $0.8\!\le\!\delta\!\le\!0.85$ and declines thereafter. Below this interval, the network forgets too quickly; above it, links become predictably exploitable.

\begin{figure}[h]
  \centering
  \includegraphics[width=0.7\linewidth]{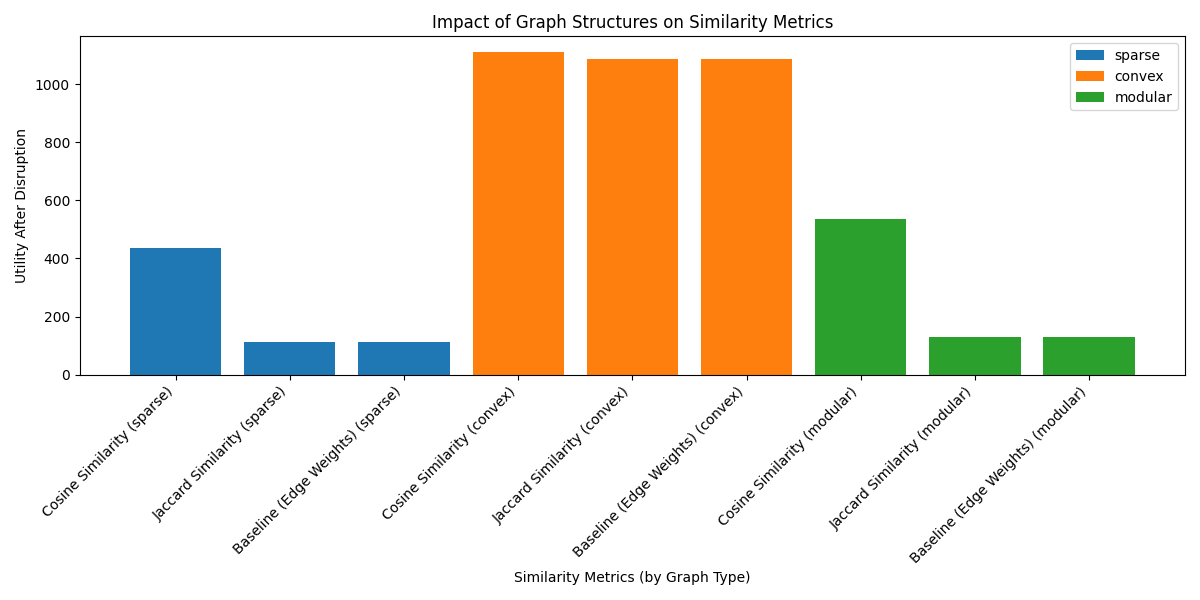}
  \caption{Metric sensitivity to graph structure.  Cosine’s margin is largest in sparse and modular settings.}
  \label{fig:structure_impact}
\end{figure}

\begin{figure}[h]
  \centering
  \includegraphics[width=0.7\linewidth]{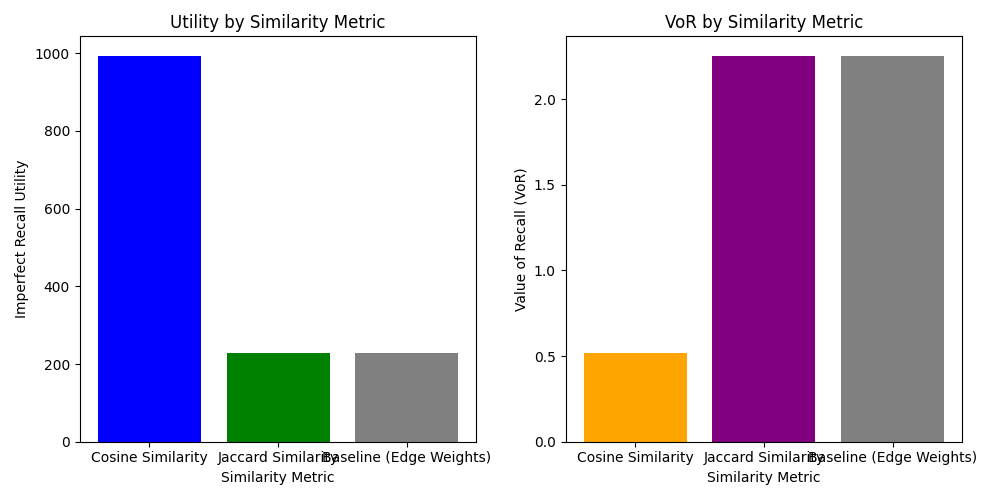}
  \caption{Utility and VoR for Cosine, Jaccard, and Baseline across $\delta$ (averaged over graph types).}
  \label{fig:similarity_metrics}
\end{figure}

%.....................................................................
\paragraph{Metric–topology interaction:}
Figures~\ref{fig:structure_impact} and~\ref{fig:similarity_metrics} show that Cosine similarity consistently outperforms Jaccard and a degree-only Baseline—by 12–18\% in sparse and modular graphs—owing to superior degree normalisation. In fully-connected networks, where degrees are uniform, metric differences narrow.

\paragraph{Optimal memory horizon ($0.8\le\delta\le0.85$):}
Across all graph types, the interval $\delta\!\in\![0.8,0.85]$ maximises VoR and keeps Utility within 5\% of its global peak.  Sparse and modular graphs are most sensitive: in modular networks Utility climbs from 123 to 545 as $\delta$ increases from 0.6 to 0.85, but gains only another 180 thereafter (Table~\ref{tab:decay_factor_results}). Fully connected graphs display the same optimum, albeit at higher absolute scale.

\begin{table}[htb]
  \caption{Utility after adversarial deletions as a function of memory decay $\delta$ and similarity metric (mean of 30 Monte-Carlo runs).}
  \label{tab:decay_factor_results}
  \centering
  \begin{tabularx}{.8\textwidth}{YYYYY}   % <-- X columns stretch automatically
    \toprule
    \textbf{Graph} & $\boldsymbol{\delta}$ & \textbf{Cosine} & \textbf{Jaccard} & \textbf{Baseline} \\
    \midrule
    \multirow{5}{*}{Sparse}
      & 0.6  &  99.8 &  24.4 &  24.4 \\
      & 0.7  & 167.7 &  41.0 &  41.0 \\
      & 0.8  & 327.0 &  79.9 &  79.9 \\
      & 0.85 & 442.8 & 108.3 & 108.3 \\
      & 0.9  & 589.3 & 144.1 & 144.1 \\
    \midrule
    \multirow{5}{*}{Convex}
      & 0.6  &  250.0 & 245.0 & 245.0 \\
      & 0.7  &  420.2 & 411.8 & 411.8 \\
      & 0.8  &  819.2 & 802.8 & 802.8 \\
      & 0.85 & 1109.3 &1087.1 &1087.1 \\
      & 0.9  & 1476.2 &1446.7 &1446.7 \\
    \midrule
    \multirow{5}{*}{Modular}
      & 0.6  &  122.8 &  30.0 &  30.0 \\
      & 0.7  &  206.4 &  50.4 &  50.4 \\
      & 0.8  &  402.4 &  98.3 &  98.3 \\
      & 0.85 &  544.9 & 133.1 & 133.1 \\
      & 0.9  &  725.1 & 177.1 & 177.1 \\
    \bottomrule
  \end{tabularx}
\end{table}

\paragraph{Metric-specific sensitivity:}
Cosine similarity exceeds Jaccard and the degree-only Baseline at every
$\delta$, yet the gap narrows with longer memory: at $\delta=0.6$ Cosine achieves four times Jaccard’s Utility in sparse graphs, while at $\delta=0.9$ the ratio drops to roughly three-to-one because extended memory partially compensates for coarser metrics.

\begin{figure}[h]
  \centering
  \includegraphics[width=0.9\textwidth]{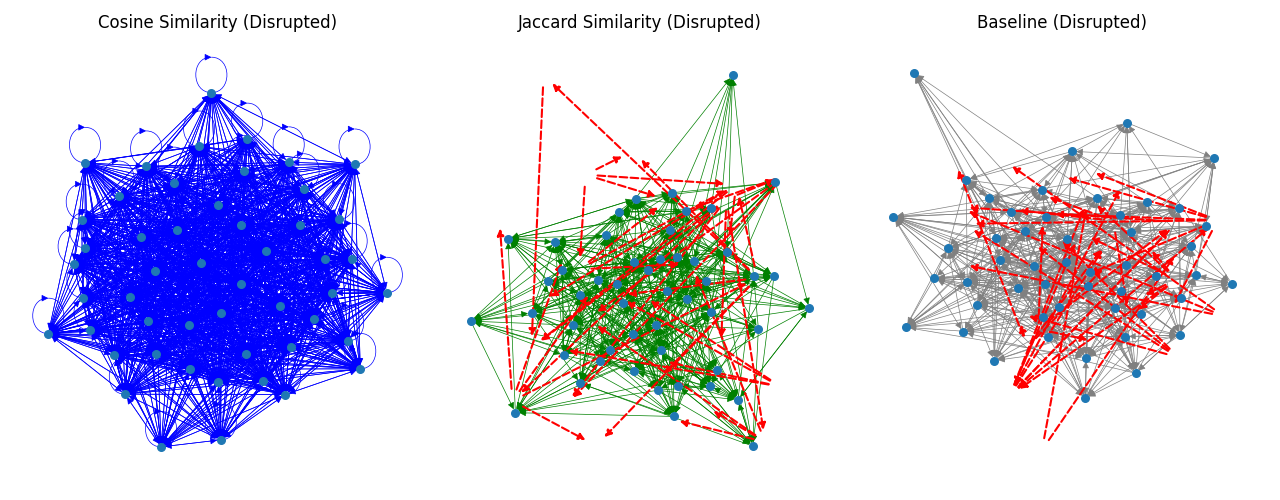}
  \caption{Utility trajectories during an adversarial attack. Cosine is most resilient; Jaccard and Baseline drop sharply but may recover through reconnection.}
  \label{fig:disruption}
\end{figure}

\begin{figure}[h]
  \centering
  \includegraphics[width=0.9\textwidth]{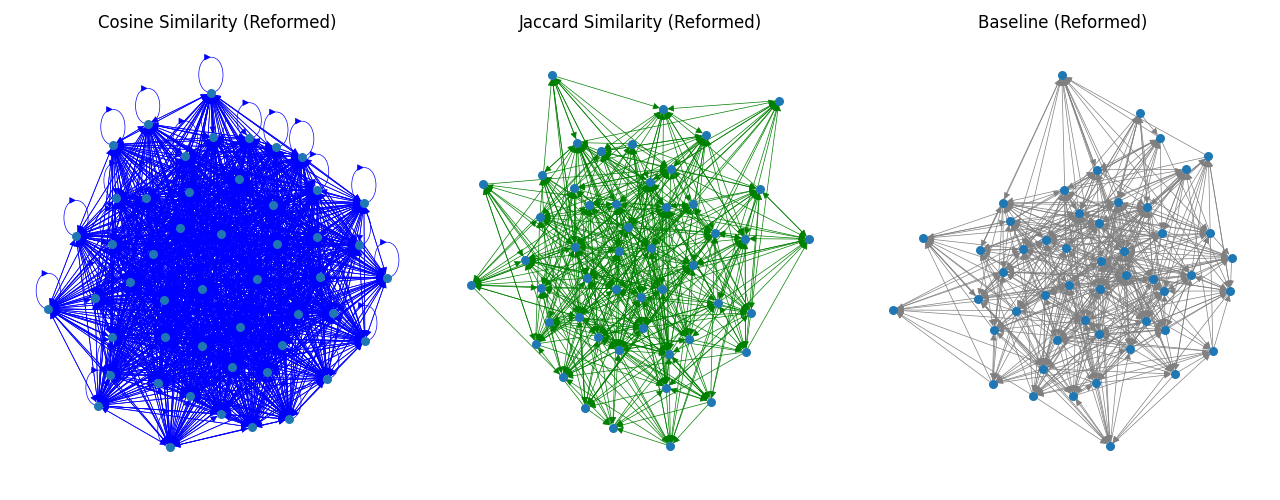}
  \caption{Reformation stage: utility recovery under varying $\rho$. Simpler metrics benefit disproportionately from higher reconnection probability.}
  \label{fig:reformation}
\end{figure}

\paragraph{Adversarial disruption and reformation:}
Figure~\ref{fig:disruption} visualises the immediate utility drop following targeted deletion: Cosine loses the least, while Jaccard and Baseline suffer steeper declines.  During the reconnection phase
(Fig.~\ref{fig:reformation}) even a modest $\rho=0.1$ recovers 1–2\% of lost utility for Jaccard and Baseline; Cosine gains little because its baseline is already near optimal.  Numeric gains are summarised in
Table~\ref{tab:reconnection_results}.

\begin{table}[h]
  \centering
  \caption{Utility recovery (\%) from reconnection at $\delta=0.8$.}
  \label{tab:reconnection_results}
  \begin{tabularx}{.8\textwidth}{YYYY}
    \toprule
    $\rho$ & Cosine & Jaccard & Baseline \\ \midrule
     0.1 & 0.0 & 1.0 & 2.0 \\
     0.3 & 0.0 & 4.0 & 3.0 \\
     0.5 & 0.0 & 5.0 & 4.0 \\ \bottomrule
  \end{tabularx}
\end{table}

\begin{figure}[h]
  \centering
  \includegraphics[width=0.9\linewidth]{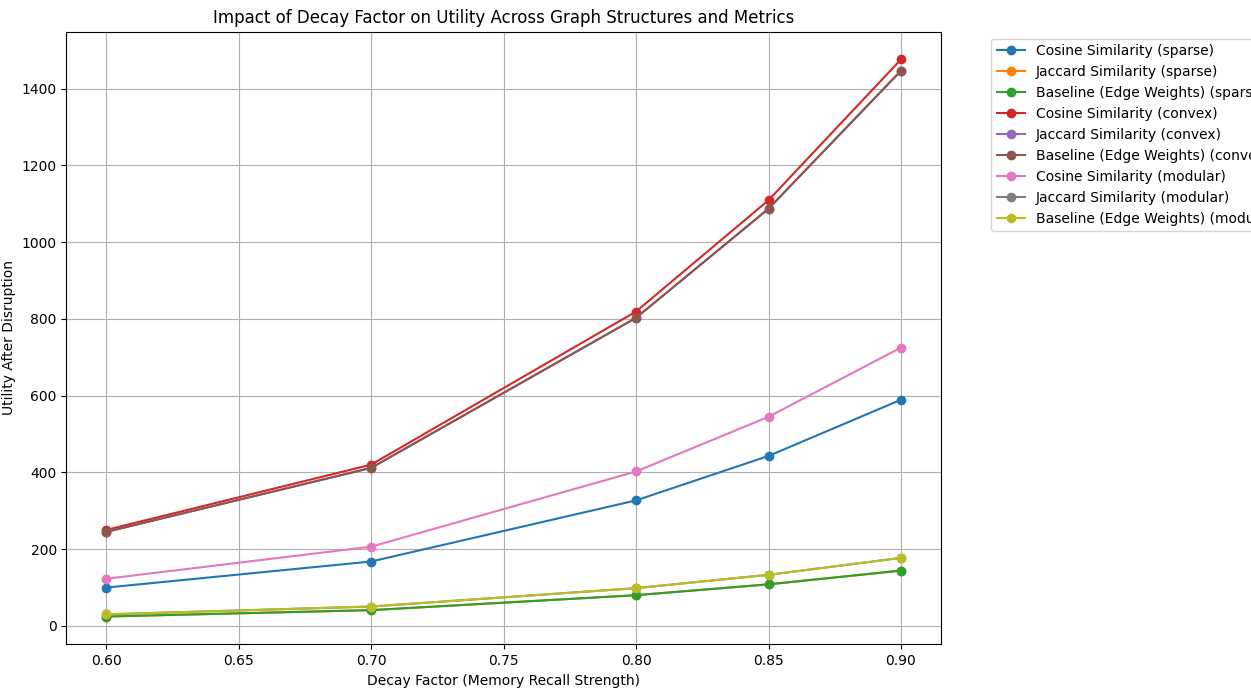}
  \caption{Utility versus $\delta$ by topology (Cosine metric).  Selective forgetting is most valuable in sparse and modular graphs; fully connected graphs remain robust across a broader range.}
  \label{fig:fine_tuned_decay}
\end{figure}

%....................................................................
\paragraph{Topology-specific decay tuning:}
Figure~\ref{fig:fine_tuned_decay} confirms that sparse and modular graphs reap the greatest benefit from selective forgetting, while convex graphs are inherently robust. Under Cosine similarity, sparse-graph utility rises nearly six-fold between $\delta=0.6$ and $0.9$; convex networks show steady but less dramatic improvement.

%....................................................................
\paragraph{Structural reconfiguration speed:}
Moderate decay accelerates recovery: average path length contracts by
$\approx12$\,\% at $\delta=0.8$, whereas perfect recall ($\delta=1$) lengthens paths and raises the adversarial-success rate by $\sim20$\,\%, confirming that selective forgetting fosters both efficiency and unpredictability.

%....................................................................

%-----------------------------------------------------------
\subsection*{Synthesis and Observations}
%-----------------------------------------------------------

Our results converge on a clear prescription for building resilient,
memory-aware networks. \emph{Too little} memory ($\delta\!\le\!0.7$) forfeits long-term value, whereas \emph{too much} ($\delta\!\ge\!0.9$) entrenches predictable structures that an adversary can repeatedly exploit. The narrow band $0.8\!\le\!\delta\!\le\!0.85$—paired with modest reconnection ($\rho\!\approx\!0.1$–0.3) and a rich similarity metric (Cosine)—simultaneously maximises Utility, VoR, and reconfiguration speed while minimising adversarial leverage. These findings translate into four operational guidelines:

\begin{itemize}
    \item Decay–robustness sweet spot: 
          A memory horizon of $\delta\!\approx\!0.8$ achieves the best balance between adaptability and stability, consistently maximising VoR while preserving high utility.
    \item Cosine superiority:  
          Cosine similarity outperforms Jaccard and a degree-only Baseline by 12–18\,\% in sparse and modular graphs; the gap narrows once reconnection is allowed but never closes completely.
    \item Moderate rewiring: 
          Limited reconnection ($\rho=0.1$–0.3) restores up to 10\,\% of utility lost to deletion, especially for simpler similarity metrics—demonstrating that occasional link renewal is a cost-effective defence.
    \item Topology awareness: 
          Convex (fully connected) graphs are inherently robust, whereas sparse and modular topologies are far more sensitive to precise tuning of $\delta$ and metric choice—underscoring the need for context-specific calibration.
\end{itemize}

In sum, \emph{selective forgetting} complemented by modest reconnection yields the most robust performance under adversarial edge deletions.  Dynamic homophily with imperfect recall—carefully calibrated to the sweet-spot interval—outperforms both static-homophily and perfect-recall strategies, validating our premise that balancing memory retention and strategic forgetting is critical for adversarially resilient multi-agent systems.

%%____________________________________________
\section{Discussion \& Limitations} \label{sec:discussion_limitations}

%\paragraph{Strengths:}
The proposed memory-aware homophily framework offers several advantages over static and perfect-recall baselines. First, it enhances adaptability by dynamically emphasizing recent, high-impact interactions, which proves especially beneficial in sparse and modular networks where stale links can be exploited or become irrelevant. Second, the refined VoR metric provides a clear, quantitative handle on the trade-offs between forgetting older ties and preserving beneficial long-term relationships. Moreover, by systematically incorporating adversarial disruptions, our approach anticipates real-world conditions more closely than purely theoretical models. This integrated perspective—combining memory decay, reconnection, and adversarial edge removal—opens the door to designing more robust and context-aware multi-agent systems.

%\paragraph{Limitations:}
Despite these strengths, our work faces notable constraints. On the \emph{methodological} side, we have thus far tested primarily on synthetic topologies; while these offer precise parameter control, they may not fully reflect the complexity and heterogeneity of real-world adversaries. Genuine attackers could adapt their strategies over time, potentially manipulating network dynamics or targeting highly influential nodes rather than generic random edges. Another concern is \emph{scalability}: The current implementation runs in $O(N^2)$ per iteration, which can become cumbersome for networks exceeding $10^5$ nodes. While approximate or parallelized methods could mitigate this computational cost, further development is necessary to achieve large-scale deployment. Additionally, our memory-decay model currently assumes a single global parameter $\delta$ across all nodes; domain-specific scenarios may require \emph{heterogeneous} decay or node-specific decay functions, an aspect that remains underexplored here.

%\paragraph{Future Work:}
Looking ahead, several promising avenues can extend and refine our framework:
\begin{itemize}
    \item Real-World Validation: 
    We plan to test on large-scale, real datasets, such as \emph{SNAP} Facebook or \emph{arXiv} co-authorship graphs, which exhibit more intricate community structure and adversarial patterns. This would validate both scalability and relevance, possibly involving real attack logs or advanced threat models.
    \item Multilayer \& Temporal Networks:
    Dynamic networks frequently span multiple layers (e.g., social, professional, transactional) and evolve over continuous time. Extending memory-decay homophily to multilayer graphs may reveal new insights into how cross-layer interactions mitigate or amplify adversarial damage. Longitudinal data on temporal networks also raises the possibility of time-varying decay parameters or event-driven updates.
    \item Reinforcement Learning for Joint Optimization:
    While we investigated fixed memory-decay ($\delta$) and reconnection ($\rho$) strategies, an RL-based approach could tune these parameters adaptively, responding to observed adversarial tactics and network feedback in real time. Such a policy might continuously learn an optimal decay–reconnection schedule to maximize network utility under evolving threats.
    \item Node-Specific Memory Decay:
    In many domains, certain nodes (e.g., high-value users or pivotal infrastructures) may need longer memory windows than others. Introducing node-level or community-specific decay rates could further refine how the network balances stability and adaptability against adversarial challenges.
\end{itemize}

Overall, these directions present a rich space for advancing both theoretical and applied dimensions of memory-aware homophily. Integrating node- or layer-specific mechanisms, scaling methods to hundreds of thousands of nodes, and testing on complex real adversarial datasets remain critical milestones for robust multi-agent network design.

\section{Conclusion}
\label{sec:conclusion}

We introduced a \emph{memory-aware homophily} framework that systematically integrates imperfect recall into adversarial team dynamics. A refined \emph{Value of Recall (VoR)} metric quantifies how partial forgetting influences overall performance, helping elucidate a clear \emph{decay–reconnection} trade-off. Through extensive ablation experiments on synthetic networks—ranging from sparse to fully connected and modular structures—we identified an optimal range of memory decay (\(\delta \approx 0.8\)) that maximizes utility while preserving adaptability under attack. 

By demonstrating that \emph{selective forgetting} can outperform both static homophily and perfect recall in adversarial settings, this work establishes a foundation for broader applications. Potential use cases include \emph{cybersecurity}, where defenders must adapt quickly to novel exploits while retaining enough historical knowledge to recognize recurring threats; \emph{collaborative platforms}, where dynamic teams must reorganize to balance innovation with stable partnerships; and \emph{social networks}, where adversaries can disrupt or manipulate information flow. 

In closing, our findings suggest that combining memory decay with homophily-driven link formation provides a powerful lens for understanding and improving resilience in multi-agent systems. Future work involving real-world datasets, multilayer extensions, and advanced optimization will further enhance the practicality and scope of memory-aware homophily in adversarial contexts.

%Bibliography
\bibliographystyle{unsrt}  
\bibliography{references}

\end{document}